\documentclass[showpacs,prl,twocolumn,aps,superscriptaddress,preprintnumbers]{revtex4}

\usepackage[dvipdfmx]{graphicx}
\usepackage{amsmath,amssymb}
\usepackage{epsfig}
\usepackage{amsmath}
\usepackage{amsfonts}
\usepackage{epstopdf}
\usepackage{color}

\usepackage{ulem}
\newcommand{\be}{\begin{equation}}
\newcommand{\ee}{\end{equation}}
\newcommand{\bear}{\begin{eqnarray}}
\newcommand{\eear}{\end{eqnarray}}

\newcommand{\ba}{\begin{array}}
\newcommand{\ea}{\end{array}}

\begin{document}
\preprint{OU-HET-896, KUNS-2628}

\title{Chaos  in chiral condensates in gauge theories}

\author{Koji Hashimoto}
\affiliation{Department of Physics, Osaka University, Toyonaka, Osaka 560-0043, Japan}
\author{Keiju Murata}
\affiliation{Keio University, 4-1-1 Hiyoshi, Yokohama 223-8521, Japan}
\author{Kentaroh Yoshida}
\affiliation{Department of Physics, Kyoto University, Kyoto 606-8502, Japan}
\date{\today}

\begin{abstract} 
Assigning a chaos index for dynamics of 
generic quantum field theories is a challenging problem,
because the notion of a Lyapunov exponent, which is useful for 
singling out chaotic behavior, works only in classical systems.
We address
the issue by using the AdS/CFT correspondence, as the large $N_c$ 
limit provides a classicalization (other than the standard $\hbar \to 0$) while keeping
 nontrivial quantum condensation.
We demonstrate the chaos
 in the dynamics of quantum gauge theories:
 The time evolution of homogeneous 
quark condensates $\langle \bar{q}q\rangle$ and $\langle \bar{q} \gamma_5 q\rangle$
in an ${\cal N}=2$ supersymmetric  QCD with the $SU(N_c)$  gauge group at large $N_c$ and 
 at large 't Hooft coupling $\lambda \equiv N_c g_{\rm YM}^2$ exhibits 
a positive Lyapunov exponent.
The chaos dominates the phase space for energy density
 $E \gtrsim (6\times 10^2)\times m_q^4(N_c/\lambda^2) $ where $m_q$ is the quark mass.
 We evaluate 
the 
largest Lyapunov exponent as a function of $(N_c,\lambda,E)$ and find that
the ${\cal N}=2$ supersymmetric QCD is more chaotic for smaller $N_c$.
\end{abstract}


\maketitle

\setcounter{footnote}{0}

{\it Introduction.}---
Revealing a hidden relation between generic 
quantum field theories and chaos is a long-standing problem.    
The
solution can ignite novel quantitative study of the complexity
of particle physics.
The problem is how one can define {a quantity like} a Lyapunov exponent, which measures
chaos, in generic quantum dynamics of field theories.
The Lyapunov exponent can be defined only in classical systems
--- once the systems are quantized, because the strong dependence on 
initial values is lost due to the quantum effect.
Then how can one measure the chaos of purely quantum phenomena,
such as the chiral condensate of QCD?

In the history {concerned with this issue}, 
the chaos of the classical limit of the Yang-Mills theory was first found
\cite{Matinyan:1981dj,Matinyan:1981ys,Savvidy:1982jk,
Biro:1994bi,Muller:1992iw,Gong:1992yu}, and 
was applied to 
an entropy production process of heavy ion collisions
\cite{Kunihiro:2008gv,Kunihiro:2010tg,Muller:2011ra,Iida:2013qwa,Tsukiji:2015rra,Tsukiji:2016krj} 
together with a color glass condensate \cite{McLerran:1993ni,McLerran:1993ka,McLerran:1994vd}.
However, the produced quark gluon plasma is strongly coupled, and a transition from the classical Yang-Mills
to quantum states is yet an open question.
On the other hand, the recent study \cite{Maldacena:2015waa} 
of out-of-time-ordered correlators of quantum fields \cite{Larkin}
defined a quantum analog
of the Lyapunov exponent, and 
opened a new direction about the problem \cite{Polchinski:2015cea,Hartman:2015lfa,Brown:2015bva,Berenstein:2015yxu,Hosur:2015ylk,Gur-Ari:2015rcq,Stanford:2015owe,Fitzpatrick:2016thx,Perlmutter:2016pkf,Turiaci:2016cvo}.
The problem of chaos in quantum dynamics could be addressed along
the line of this development.

%

We here provide a solution of the problem. 
A key
observation is that there exist several ways to relate quantum field theories to classical ones, although the standard
method is the semiclassical limit $\hbar  \to 0$.
In fact, 
{a} large $N$ limit of strongly coupled gauge theories is another classical limit. 
We use the AdS/CFT correspondence 
\cite{Maldacena:1997re} as a {tool to resolve}
the problem and to map the strongly coupled theories to a
classical gravity, which enables us to calculate the Lyapunov exponents of expectation values of
operators directly probing the quantum dynamics.
The idea is supported by
recent analyses of chaotic motion of classical strings in AdS-like spacetimes
\cite{Zayas:2010fs,Basu:2011di,Basu:2011dg,Basu:2011fw,Basu:2012ae,Ma:2014aha,Bai:2014wpa,Ishii:2015wua,Asano:2015qwa} (see also \cite{Aref'eva:1997es,PandoZayas:2012ig,Basu:2013uva,Giataganas:2013dha,Farahi:2014lta,Asano:2015eha}).

In this Letter we first show that the linear $\sigma$ model of low-energy QCD exhibits chaos of
the chiral condensate, which serves as a toy model of chaos of a quantum phenomenon.
Then we concretely study an ${\cal N}=2$ supersymmetric QCD with the $SU(N_c)$ ${\cal N}=2$ 
gauge group at large $N_c$ and at strong coupling
\cite{Karch:2002sh}. By using the AdS/CFT, we 
calculate Lyapunov exponents 
of the time evolution of a homogeneous quark condensate.
The analysis shows how the complexity of the quantum dynamics depends on $N_c$ and $\lambda$: 
The theory is more chaotic for a larger 
$\lambda$ or a smaller $N_c$.
The discovered chaos is a quantum analog
of the butterfly effect. 
We discuss that our Lyapunov exponent is also described by a generalized out-of-time-ordered correlator.


\vspace{5mm}
{\it Chaos in a linear $\sigma$ model.}---
The most popular 
effective action for the chiral condensate of 
QCD is the linear $\sigma$ model. It describes a universal class of theories
governed by a chiral symmetry via a spontaneous and an explicit
breaking. The former comes from the QCD strong coupling dynamics
while the latter comes from a quark mass term. 
We find below that the model exhibits chaos.

The simplest linear $\sigma$ model is with a chiral $U(1)$ symmetry
with an explicit breaking (quark mass) term: 
\begin{align}
&S=\int\!d^4x
\left\{
-\frac12 [(\partial_\mu\sigma)^2+(\partial_\mu\pi)^2]-V
\right\},
\label{lsm}
\\
&
V\equiv\frac{\mu^2}{2}(\sigma^2+\pi^2)
+\frac{g_4}{4}(\sigma^2+\pi^2)^2+a \sigma + V_0.
\end{align}
Here for simplicity we consider only a single flavor and ignore
the axial anomaly. $\sigma(x^\mu)$ and $\pi(x^\mu)$ are fields whose
fluctuation provides a sigma meson field with the mass 
$m_\sigma$ and a neutral pion field with the mass
$m_\pi$, respectively. The vacuum expectation value of $\sigma$
minimizing the potential $V$ defines the chiral condensate: $\langle\sigma\rangle =f_\pi$. 
A constant $V_0$ is introduced just for shifting the vacuum energy to zero.
Relations to observable parameters are found as 
$2\mu^2=-m_\sigma^2+3m_\pi^2$, $g_4 = (m_\sigma^2-m_\pi^2)/(2f_\pi^2)$,
and $a=-m_\pi^2f_\pi$\footnote{
The action can also be thought of as a neutral part of a non-Abelian linear $\sigma$ model
with three pions. 
}.

\begin{figure}
\includegraphics[width=4cm]{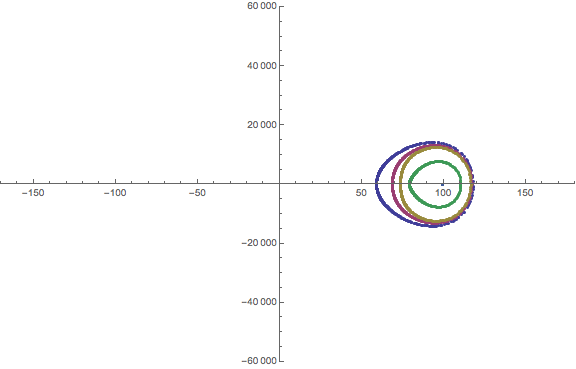}
\includegraphics[width=4cm]{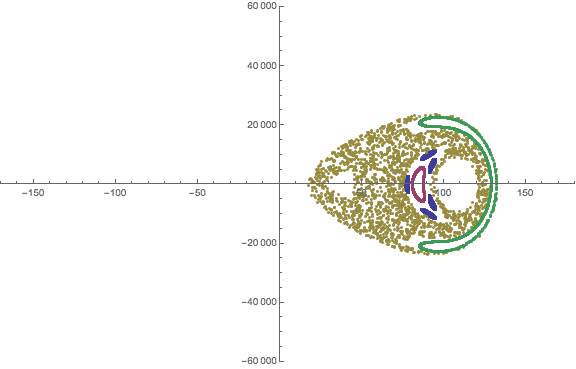}
\includegraphics[width=4cm]{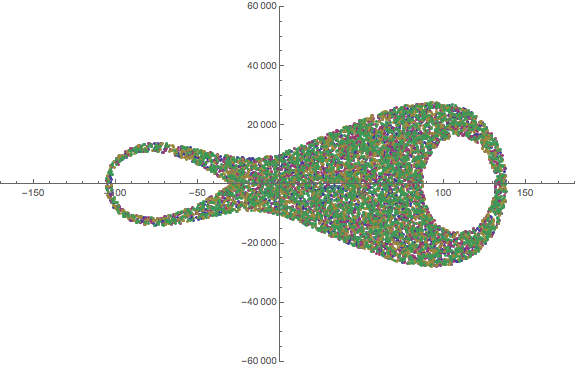}
\includegraphics[width=4cm]{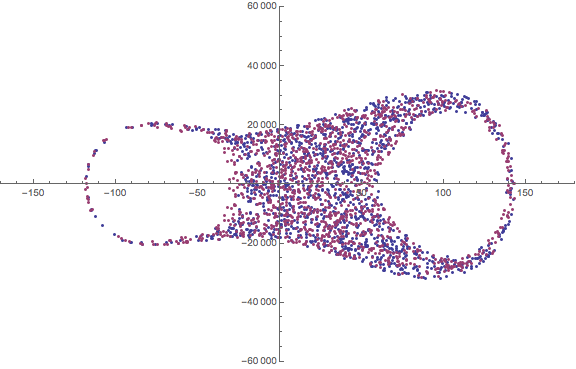}
\includegraphics[width=4cm]{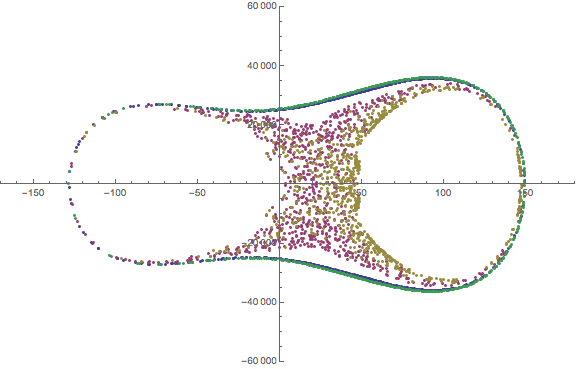}
\includegraphics[width=4cm]{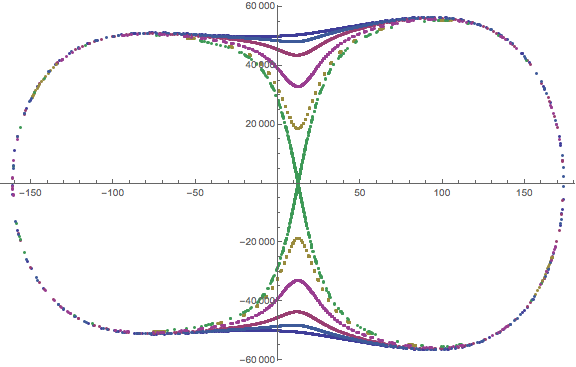}
\caption{The Poincar\'e sections 
for the linear sigma model.
The horizontal axis is $\sigma(t)$, while the vertical axis is $\dot{\sigma}(t)$. 
The section is chosen by $\pi(t)=0$.
The energy density is chosen as
$E^{1/4} =$ 100, 130, 140, 150, 160 and 200[MeV] 
in the 
top-left, top-right, middle-left, middle-right, lower-left, and lower-right 
figures, respectively. 
}
\label{fig:sigma}
\end{figure}

Let us consider a homogeneous motion \footnote{The homogeneity is an assumption, but
there are examples of successful homogeneous study such as chiral random matrix theory. 
} 
of the $\sigma$ model fields $\sigma(t)$ and $\pi(t)$,
which is a time evolution of the chiral condensate. In a Hamiltonian language,
there are four dynamical variables $(\sigma,\pi,\dot{\sigma},\dot{\pi})$
while the conserved quantity is only the total energy, so there may exist chaos.
We search the chaos by varying the total energy density $E$, and find a chaotic behavior
of the chiral condensate. At an intermediate scale of the energy density, the Poincar\'e
section 
{ 
(a cross section of orbits in the phase space with sampled initial conditions sharing
a chosen conserved energy)
}
exhibits a 
scattered plot, which is chaos; see Fig.~\ref{fig:sigma}. For the numerical simulations,
we have chosen $m_\pi= 135$[MeV], $m_\sigma= 500$[MeV], and $f_\pi=93$[MeV].

\begin{figure}
\includegraphics[width=6cm]{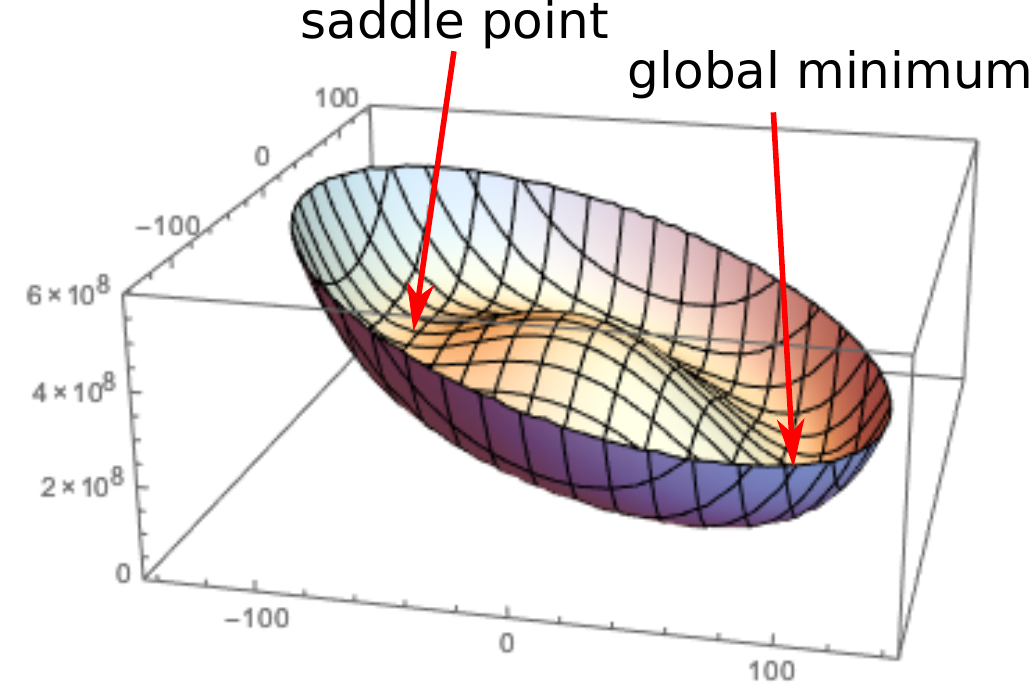}
\caption{The potential $V$ of the linear $\sigma$ model. The horizontal axes are
for $\sigma$ and $\pi$. The potential bottom is at $(\langle\sigma\rangle,
\langle\pi\rangle)=(f_\pi,0)$.
Because of the quark mass term, there appears a separatrix 
on the negative axis of $\sigma$.}
\label{fig:potsigma}
\end{figure}

{ 
In this model, the chaos emerges consequently due to 
the existence of a saddle point in the potential $V$ as shown in Fig.~\ref{fig:potsigma}.
In general, 
separatrices (boundaries between phase space domains with distinct dynamical behavior)
are associated with saddle points. They
are broken under weak perturbations 
and become a seed of chaos.
{The separatrix in the potential $V$ is generated by} 
the combination of the explicit and the spontaneous symmetry
breaking terms.  
For example, a potential with
no $a \sigma$ term makes the system  
}
integrable
due to the Poincar\'e-Bendixon theorem.

\begin{figure}
\includegraphics[width=6cm]{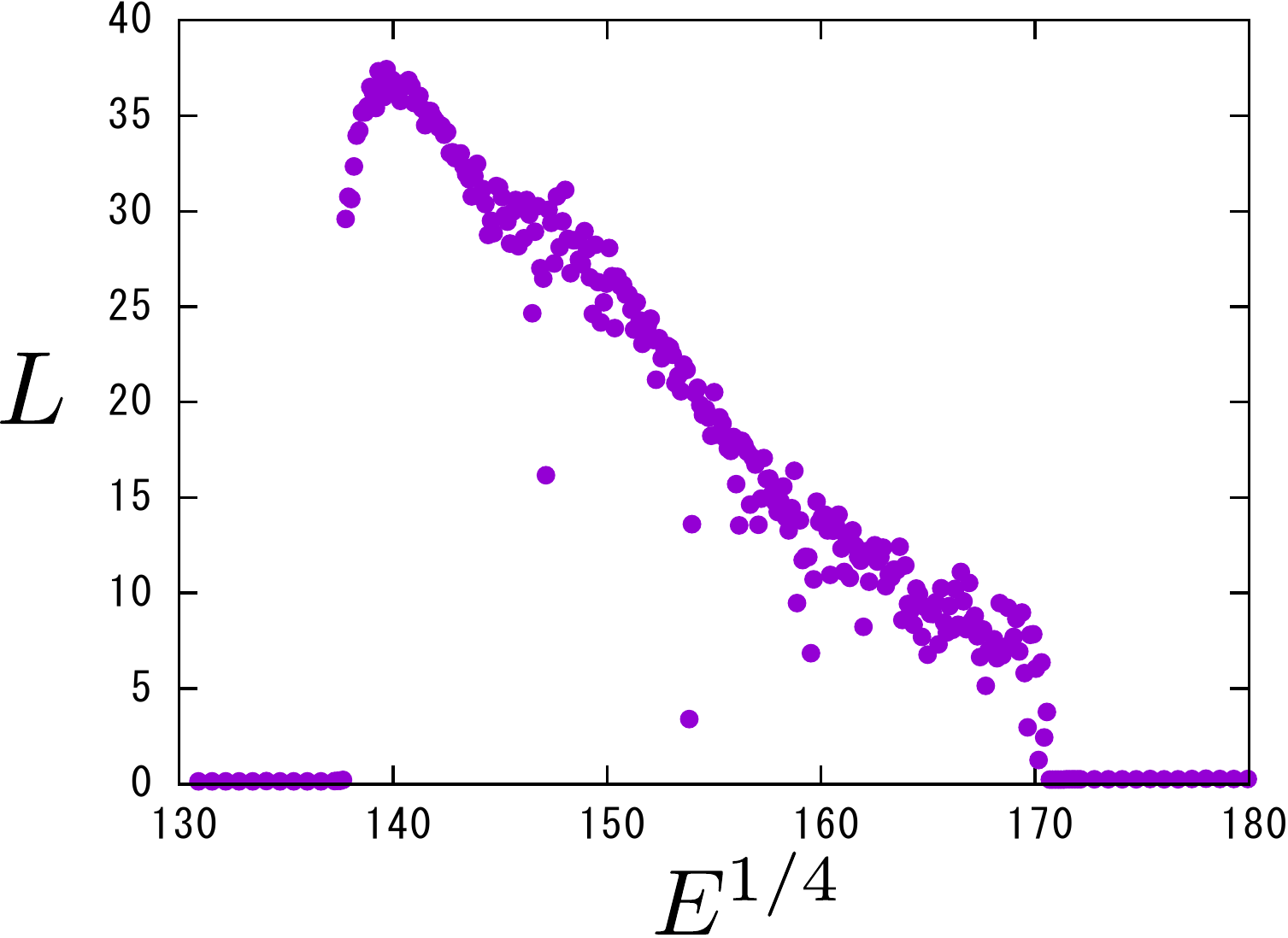}
\caption{The Lyapunov exponent $L$ [MeV] 
of the linear $\sigma$ model as a function of the energy density $E^{1/4}$ [MeV]. 
The initial condition is chosen as $\sigma=f_\pi$ and 
$\dot{\sigma}=\dot{\pi}=0$.}
\label{fig:Lsigma}
\end{figure}

It is interesting that the chaotic phase (at which the Poincar\'e section is covered mostly by ergodic
chaos pattern) appears only at an intermediate scale of the energy density: 
$1.3\times 10^2$[MeV] $<E^{1/4}< 1.7\times 10^2$[MeV]. 
It is roughly equal to the height of the separatrix $\sim m_\pi^2f_\pi^2$. 
The measure of the chaos is provided by the Lyapunov exponent
\begin{align}
L(E) \equiv \lim_{t\to\infty}\lim_{d(0)\to 0} \frac{1}{t}\log 
\frac{d_{\langle\bar{q}q\rangle}(t)}{d_{\langle\bar{q}q\rangle}(0)}
\label{Lyapdef}
\end{align}
where $d_{\langle\bar{q}q\rangle}(t)$ is the distance between the two time evolution
orbits of the quark condensate 
$\sigma(t)$ 
{ 
and $d_{\langle\bar{q}q\rangle}(0)$ is taken to be infinitesimally small.
}
The energy dependence of the calculated Lyapunov exponent of the linear $\sigma$ model
is given in Fig.~\ref{fig:Lsigma}. We observe that chaos appears only at the intermediate energy scale.
It suggests that the thermal entropy of QCD might be 
related to the Lyapunov exponent and to an entropy production of the thermal history of
the Universe in some manner.

\vspace{5mm}
{\it Chaotic chiral condensate.}---
Our analysis suggests 
that 
generically chaos appears in the time evolution of chiral condensates, because the linear $\sigma$
model is just based on a symmetry and its breaking. 
{ 
Although there are various $\sigma$ models of QCD, they contain the simplest linear $\sigma$ model
\eqref{lsm} as a subsector. Generic $\sigma$ models concern non-Abelian chiral symmetries
$U(N_f)$ with $N_f$ quark flavors, and the}
hidden local symmetry \cite{Bando:1984ej,Bando:1987br} 
can be used to formulate vector meson actions. Generically
non-Abelianization accompanies a specific nonlinearity due to the hidden symmetry, which
is another possible nest of chaos. 

Unfortunately,  linear or non-linear $\sigma$ models 
are toy models in which
a classical treatment is not simply justified,
and, furthermore, they describe only
a universality class, 
so a precise relation to 
QCD is lost. Only with the
large $N_c$ limit, { 
classicalization is certified and} an explicit connection is found.
In the following, we resort to the AdS/CFT
correspondence with which the large $N_c$ 
{ and the large $\lambda$ limits lead to an exact classical theory of mesons
and chiral condensates.}
Using the AdS/CFT correspondence,
a chaotic index such as the Lyapunov exponent can be calculated as a function of
theories.

\vspace{5mm}
{\it Action for the quark condensates from AdS/CFT.}---
In AdS/CFT correspondence \cite{Maldacena:1997re}, chiral condensates at
large $\lambda$ and large $N_c$ can be seen at the asymptotic behavior of
bulk fields corresponding to the gauge-invariant operators such as 
$\langle \bar{q}q\rangle$ and $\langle \bar{q} \gamma_5 q\rangle$. 
In this Letter, as a first step, we analyze the most popular holographic model,
the ${\cal N}=2$ supersymmetric QCD. 
In the theory with $N_f$ hypermultiplets of fundamental quarks coupled to
the ${\cal N}=4$ supersymmetric Yang-Mills theory, the quark sector is
introduced as $N_f$ probe D7-branes \cite{Karch:2002sh} in the geometry of
AdS$_5\times$S${}^5$
(see \cite{Erdmenger:2007cm} for a review). 
The static quark condensates vanish due to the supersymmetries.
We are interested in the time-dependent dynamics of the condensates,
which is directly encoded in the D7-brane 
action we calculate in the following. 

Any chaos needs nonlinear terms, and the D7-brane action suffices the need.
For multiflavor case $N_f\geq 2$ the action possesses a non-Abelian symmetry $U(N_f)$
and is effectively described by a massive $SU(N_f)$ Yang-Mills theory. 
There are two adjoint scalar fields which measure the fluctuation of the $N_f$ D7-brane
worldvolume in the transverse directions, and those vacuum expectation values
are the condensates $\langle \bar{q}q\rangle$ and $\langle \bar{q} \gamma_5 q\rangle$.
Let us evaluate the non-Abelian D-brane action proposed in Ref.\cite{Myers:1999ps}:
\begin{align}
S_{\rm D7} = - T_{\rm D7}\int \! d^8\xi \; {\rm STr}\sqrt{-\det \tilde{G}_{rs}}\sqrt{\det Q^a_{\; b}}
\label{acmy}
\end{align}
where $\tilde{G}_{rs} \equiv G_{rs} + G_{ra}(Q^{-1}-\delta)^{ab} G_{sb} (r,s=0,\cdots,7)$ and 
$Q^a_{\; b} \equiv \delta^a_{\;b} + i[X^a,X^c]G_{cb}/2\pi\alpha' (a,b=8,9)$ . We took a static gauge and have ignored
gauge fields on the D7-branes, and $G_{rs}\equiv g_{rs}(X)+\partial_rX^a
\partial_sX^bg_{ab}(X)$ is the induced metric on the D7-branes.
The AdS$_5\times$S${}^5$ metric $g(X)$ is
\begin{align}
ds^2 = \frac{r^2}{R^2}(dx^\mu)^2 + \frac{R^2}{r^2}
(d\rho^2+\rho^2d\Omega_3^2+(dX^8)^2+(dX^9)^2)
\nonumber
\end{align}
where $X^8$ and $X^9$ are directions transverse to the D7-brane,
and $r^2 \equiv \rho^2 + (X^8)^2+(X^9)^2$. Here
$R\equiv (2\lambda)^{1/4} (\alpha')^{1/2}$
is the AdS radius, and $T_{\rm D7}\equiv (2\pi)^{-6}(\alpha')^{-4}g_{\rm YM}^{-2}$ is the D7-brane tension.
``STr'' means a symmetrized trace in the $U(N_f)$ adjoint indices.

A static classical solution of the D7-brane action was found in Ref.\cite{Karch:2002sh} as
$(X^8,X^9)=(c,0)$ in which $c$ is related to the quark mass as $c = 2\pi\alpha' m_q$.
The D7-brane solution independent of $\rho$ means vanishing condensates
$\langle \bar{q}q\rangle=\langle \bar{q} \gamma_5 q\rangle=0$, since these expectation values
are coefficients of $1/\rho^2$ appearing in $X^8(\rho)$ and $X^9(\rho))$ at $\rho \sim \infty$. 
Fluctuations $(w^8,w^9)\equiv (X^8-c,X^9)$ correspond to towers of  scalar or pseudoscalar mesons of the 
theory, and the action quadratic in the fluctuations provides the spectra of the mesons
\cite{Kruczenski:2003be}.
A linear analysis of the action (\ref{acmy}) concerning a part of the commutator term 
was found in Ref.\cite{Erdmenger:2007vj}. We need the full structure of the commutator term. 
Expanding the action (\ref{acmy}) around the classical solution up to 
a quadratic order in $\partial X$ and also up to a single commutator term $[X,X]^2$,
we obtain
\begin{align}
S = -T_{\rm D7}\int \! \rho^3 d^4x d \rho d\Omega_3
\; {\rm STr} \left[
1+\frac{R^4 (\partial_\mu w^a)^2}{2(\rho^2+c^2)^2}
\right.
\nonumber
\\
\left.
+ \frac{(\partial_\rho w^a)^2}{2}
- \frac{R^4 [w^8,w^9]^2}{2(2\pi\alpha')^2(\rho^2+c^2)^2}
\right].
\label{wac}
\end{align}
The expansion is valid for $w^a\ll c$ and $|\partial_\mu w^a|\ll c^2/R^2$. We assumed that $w^a$
is independent of $\Omega_3$ for simplicity. 

We are interested in low-energy region, so we excite only the lowest meson eigenstate
$w^a = ({\cal N}/(\rho^2 + c^2))\phi_a(t)$ and substitute it to (\ref{wac}). The normalization
${\cal N}$ 
is fixed to have a canonical kinetic term for the lightest scalar or pseudoscalar mesons $\phi_a$. 
The resultant action for spatially homogeneous meson fields is
\begin{align}
S =  \int d^4x \, {\rm Tr}\left[
\frac12 \dot\phi_a^2 -\frac{8\pi^2 m_q^2}{\lambda} \phi_a^2 + \frac{36 \pi^2}{5 N_c}[\phi_8,\phi_9]^2
\right] .
\label{phiac}
\end{align}
The matrix elements of the expectation value of the mesons are the condensates of
the flavor $i,j(=1,\cdots,N_f)$:
\begin{align}
\left(\phi_8^{ij}(t) \, , 
\phi_9^{ij}(t)\right)\propto
\Bigl(\langle \bar{q^i}q^j(t)\rangle \, ,  \langle \bar{q^i}\gamma_5 q^j(t)\rangle\Bigr) \, .
\end{align}
{ At the static vacuum the condensates vanish in this ${\cal N}=2$ supersymmetric QCD.}

\begin{figure}
\includegraphics[width=4cm]{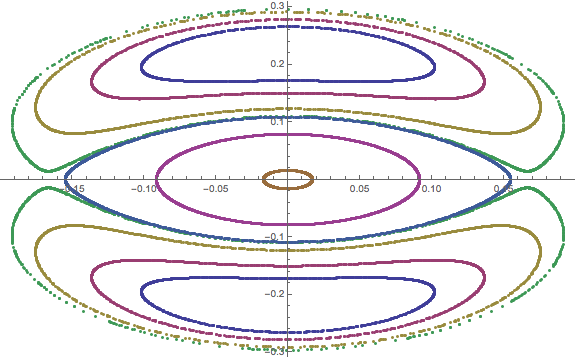}
\includegraphics[width=4cm]{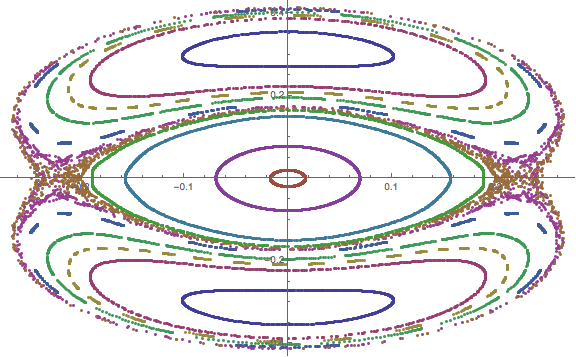}
\includegraphics[width=4cm]{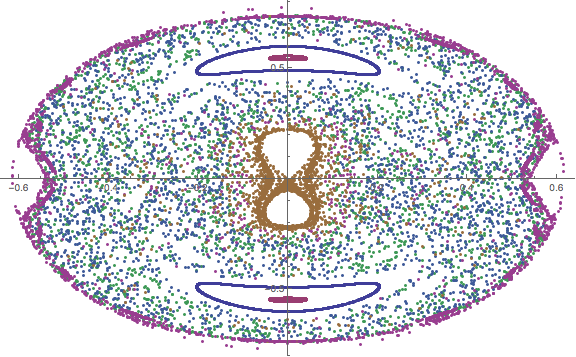}
\includegraphics[width=4cm]{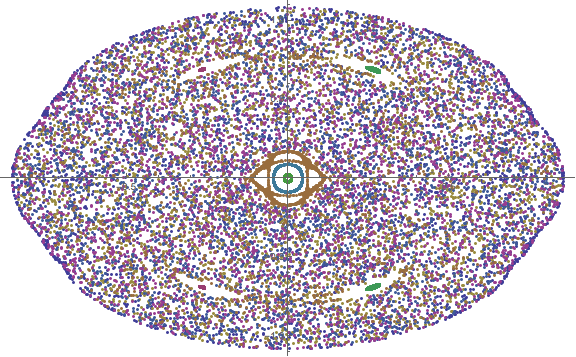}
\includegraphics[width=4cm]{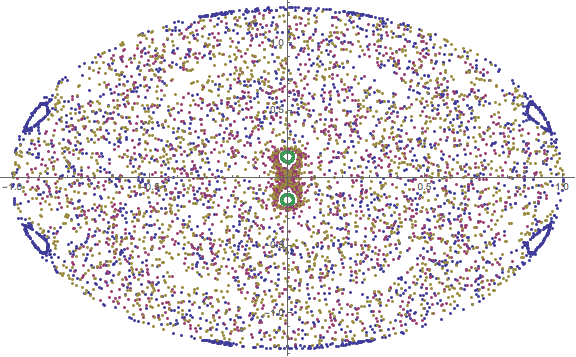}
\includegraphics[width=4cm]{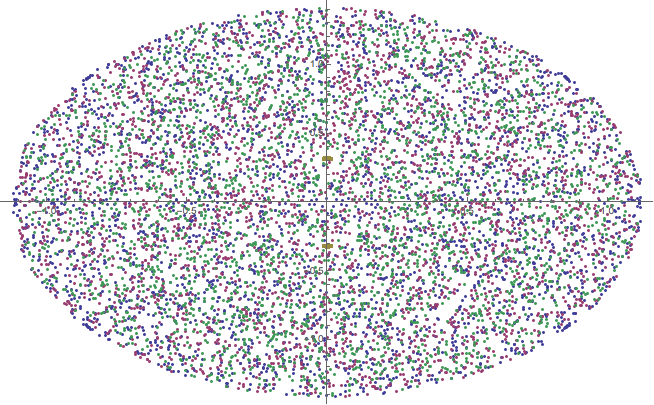}
\caption{The Poincar\'e sections for $\lambda=100$ and $N_c=10$. The horizontal axis is 
$y(t)$, while the vertical axis is $\dot{y}(t)$. The section is chosen by $x(t)=0$.
The energy is chosen as
$E = 0.05, 0.1,0.3,0.6,0.8,1$ in the top-left, top-right, middle-left, middle-right, lower-left, and lower-right figures, respectively, in the unit $m_q=1$.}
\label{fig:Poi}
\end{figure}

\vspace{5mm}
{\it Chaotic behavior.}---
To extract the simplest nonlinearity, we consider $N_f=2$ and 
a subsector $\phi_8 = x(t)\sigma_1/\sqrt{2}$ and $\phi_9 = y(t)\sigma_2/\sqrt{2}$.
This respects the equations of motion of (\ref{phiac}).
Then the system reduces to a classical mechanics of a quartic oscillator with the action
\begin{align}
S_0 = \int \! dt \, \left[\frac12 (\dot{x}^2+\dot{y}^2) - \frac{m^2}{2}(x^2+y^2) - g x^2 y^2\right].
\label{acq}
\end{align}
Here $m\equiv 4\pi m_q/\sqrt{\lambda}$ is the meson mass, and the quartic coupling is 
$g \equiv 72\pi^2/(5N_c)$.
This is a well-known model of chaos \cite{Matinyan:1981ys,Savvidy:1982jk}
(see \cite{Biro:1994bi} for a review). In Fig.\ref{fig:Poi}, we show six numerical plots of Poincar\'e sections
of the system as an example to visualize the chaos.\footnote{
The effective mass of the field $y(t)$ in Eq.(\ref{acq}) is given by $m^2-2gx^2$ and, thus, the oscillation of $x(t)$ can cause the parametric resonance. This would be regarded as the origin of the chaos in the current system.
} As the energy density increases, the system
clearly shows an order-chaos phase transition.
Indeed, it is known that, in this system (\ref{acq}),
above a certain energy  
chaos dominates the phase space
\cite{Matinyan:1981ys,Savvidy:1982jk,Salasnich:1997cw}. On the other hand, for a lower energy,
the system is in an ordered phase and the motion is regular.
So, we conclude that the time evolution of the homogeneous quark condensate of
the ${\cal N}=2$ supersymmetric QCD at strong coupling and at large $N_c$ has a chaotic phase.

Let us study the strength of the chaos as a function of the theory, $\lambda, N_c$ and $m_q$.
The system is invariant \cite{Matinyan:1981ys,Savvidy:1982jk} under the following
scaling symmetry: $x\to \alpha x$, $y \to \alpha y$, $t \to \beta t$, and $\lambda \to \beta^2 \lambda$, 
$N_c \to \alpha^2\beta^2 N_c$, $m_q\to m_q$, $E \to (\alpha^2/\beta^2) E$.
So, a scale-invariant combination $E \lambda^2/N_c$ 
governs the dynamical phase of the system. The chaos-order
phase transition occurs at a critical energy scale $E_{\rm chaos}$ 
above which the Poincar\'e section
is covered mostly by the ergodic chaos pattern. Our numerical calculation shows 
$E_{\rm chaos}\sim 0.6 m_q^4$
for $\lambda=100$ and $N_c=10$, so together with the scaling argument we obtain
\begin{align}
E_{\rm chaos} \sim (6\times 10^2) \times m_q^4 \frac{N_c}{\lambda^2}.
\label{chaosene}
\end{align}
Therefore, the energy region for chaos increases for smaller $N_c$ or larger $\lambda$.
We 
 { conclude} that our ${\cal N}=2$ supersymmetric QCD is more chaotic for 
smaller $N_c$ or larger $\lambda$.

%
%
%

\begin{figure}
\includegraphics[width=4.2cm]{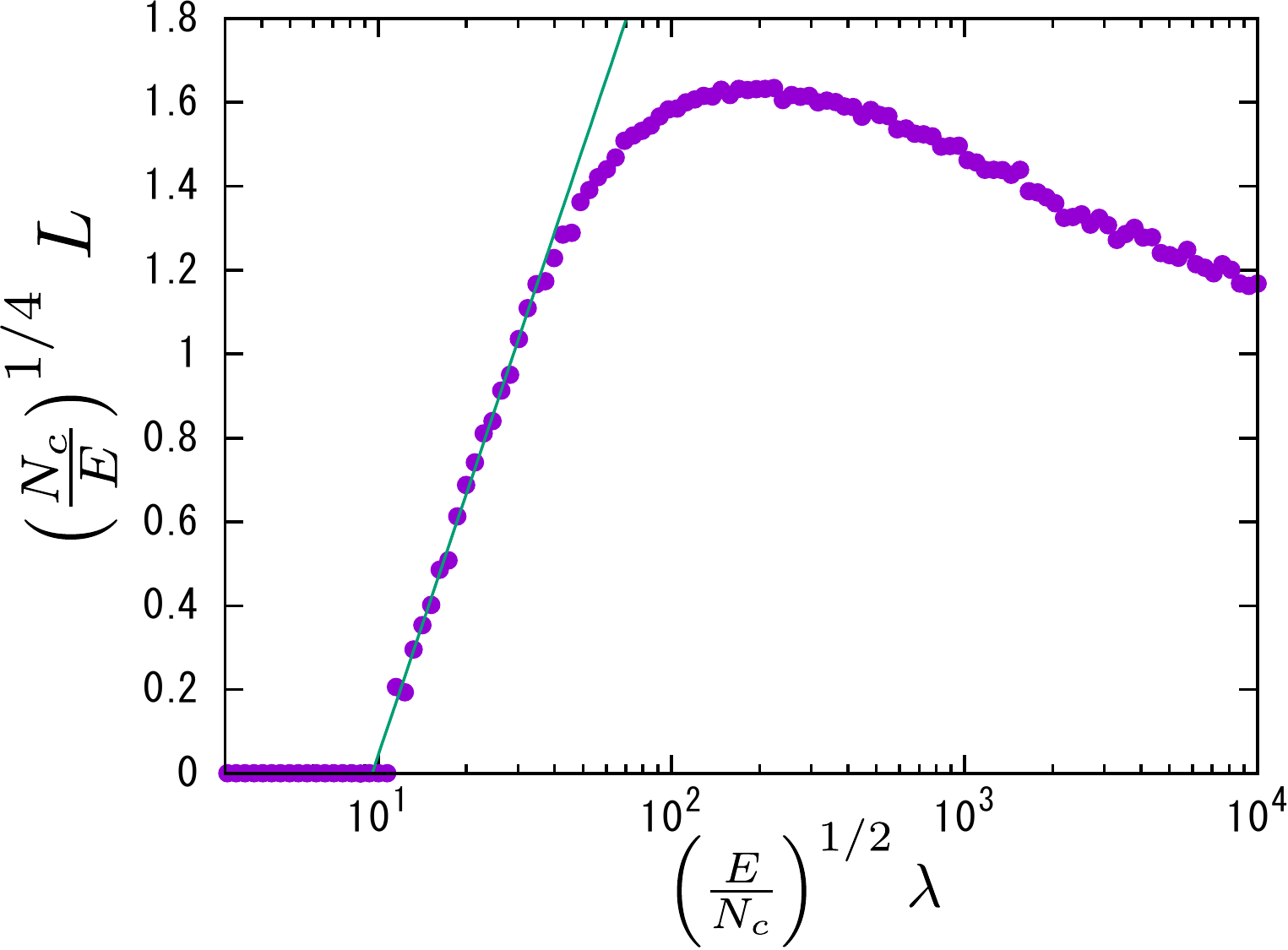}
\includegraphics[width=4.2cm]{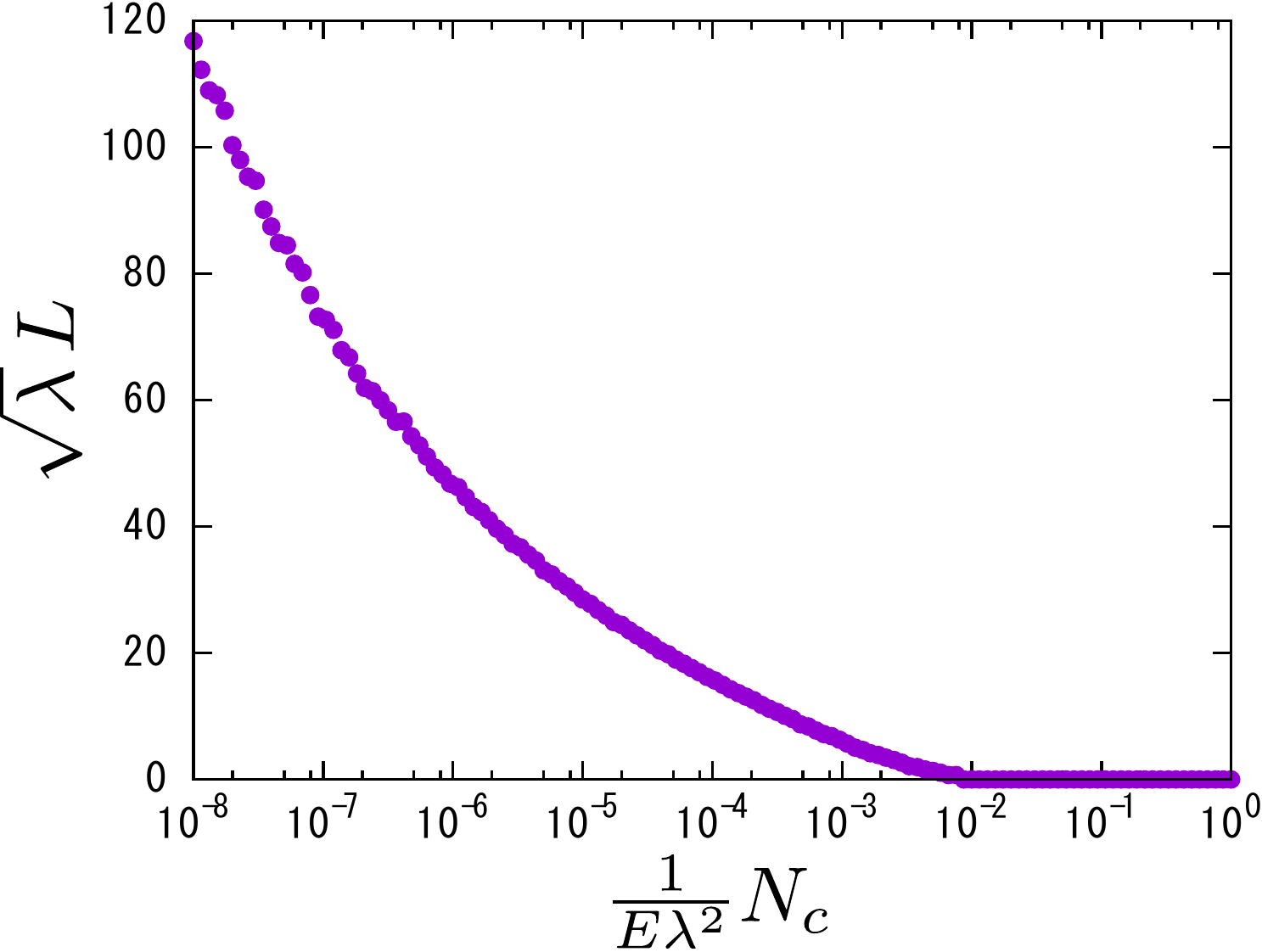}
\vspace{-3mm}
%
\caption{The Lyapunov exponent $L$ as functions of $\lambda$(left) and $N_c$(right) in the unit of $m_q=1$.
}
\label{fig:N}
\end{figure}

By the scaling transformation described above, the Lyapunov exponent, which has the dimension of inverse time, is scaled as
$L\to \beta^{-1}L$.
In Fig.~\ref{fig:N}, our numerical evaluation of the Lyapunov exponent is shown.
In Fig.~\ref{fig:N} (left), $(N_c/E)^{1/4}L$ is plotted as a function of $(E/N_c)^{1/2}\lambda$, 
where the horizontal and vertical axes are taken as scaling-invariant combinations.
This figure is convenient to see the $\lambda$ dependence of the Lyapunov exponent for fixed $N_c$ and $E$.
For $(E/N_c)^{1/2} \lambda\lesssim 10 $, one can see that there is no chaos i.e., $L=0$.
For $10 \lesssim(E/N_c)^{1/2} \lambda \lesssim 200$, the Lyapunov exponent increases linearly as a function of $\log \lambda$.
Fitting the plots in this region, the following formula is obtained: 
\begin{align}
L\simeq \left(\frac{E}{N_c}\right)^{1/4}\left[0.90\log \left\{\left(\frac{E}{N_c}\right)^{1/2}\lambda\right\} - 2.03\right] \, .
\label{Lnum}
\end{align}
Slightly above the critical energy scale $E_\textrm{chaos}$, 
the Lyapunov exponent can be approximated by this simple expression.
For large $\lambda$, it deviates from (\ref{Lnum}) and has a maximum value $L\simeq 1.6 \times (E/N_c)^{1/4}$ at $\lambda\simeq 200\times (N_c/E)^{1/2}$.
For $(E/N_c)^{1/2} \lambda\gtrsim 200$, the Lyapunov exponent decreases because the mass term disappears
and chaos is expected to be saturated by that of the pure massless Yang-Mills.

In Fig.~\ref{fig:N} (right), the same result is shown in a different normalization, $N_c/(E \lambda^2)$ vs $\lambda^{1/2}L$.
This is convenient to see $N_c$ dependence for fixed $\lambda$ and $E$.
From the figure, we can find that Lyapunov exponent is a decreasing function of $N_c$ for fixed $E$ and $\lambda$.
{ Therefore, we conclude that the strongly coupled large $N_c$ 
${\cal N}=2$ supersymmetric QCD is more chaotic for
a smaller $N_c$.}

\vspace{5mm}
{\it Outlook.}---
We {have} explicitly showed that Lyapunov exponents 
can be calculated for chiral condensates
by using the large $N_c$ limit, which amounts to solving the problem of assigning a chaos index to
quantum dynamics. 

Let us note that our Lyapunov exponent $L$ 
can be written as an out-of-time-ordered correlator
\begin{align}
e^{2L t} \sim \langle E,{\cal M}|[Q(t),P(0)]^2 |E,{\cal M}\rangle_{N\gg 1, \lambda \gg 1}
\end{align}
where $Q(t)\equiv \bar{\psi}\psi(t)$ is the chiral condensate operator inserted at time $t$,
and $P$ is for its shift, $[Q(x),P(x')]= i \delta(x-x')$. The state $|E,{\cal M}\rangle$
is an energy eigenstate of the supersymmetric QCD Hamiltonian, with
a degeneracy index ${\cal M}$.\footnote{See the
supplemental material for details, which includes Ref.~\cite{Hasenfratz:2009mp}.} The original out-of-time-ordered correlator uses a 
thermal partition \cite{Maldacena:2015waa,Larkin}, while ours is 
an energy eigenstate, so the temperature scale of the former roughly corresponds to our 
energy $E$.

Our method can assign a Lyapunov exponent to quantum dynamics of gauge theories
and opens broad applications of chaos to particle physics. Possible arenas may include
entropy production (see \cite{Latora:1999}), anarchy neutrino masses \cite{Hall:1999sn}
and Higgs criticality \cite{Buttazzo:2013uya} and related inflations 
\cite{Salvio:2013rja,Hamada:2014xka,Hamada:2014wna}. 
It would be interesting to find some relations between fundamental physics and chaos.

%

\vspace{5mm}
\begin{acknowledgments}
We have greatly benefited from the advice and comments by S.~Sasa, and are grateful to him.
We also thank valuable discussions with L.~Pando Zayas, S.~Heusler, A.~Ohnishi, K.~Fukushima, 
B.~-H.~Lee, and H.~Fukaya.
The work of K.~H.~was supported in part by JSPS KAKENHI Grant Numbers 15H03658, 15K13483.
The work of K.~M.~was supported by JSPS KAKENHI Grant Number 15K17658.
The work of K.~Y.~was supported by the Supporting Program for
Interaction-based Initiative Team Studies
(SPIRITS) from Kyoto University and by a JSPS Grant-in-Aid for Scientific
Research (C) No.\,15K05051.
This work was also supported in part by the JSPS Japan-Russia Research
Cooperative Program
and the JSPS Japan-Hungary Research Cooperative Program.
\end{acknowledgments}

\vspace{10mm}


\appendix

\section{--- Supplemental material --- \\
Out-of-time-ordered correlator and classical chaos}

In this supplemental material, we describe how 
classically measured Lyapunov exponent
of a deterministic chaos can be related to the out-of-time-ordered correlators for quantum fields.
In particular, 
our Lyapunov exponent measured in
the linear sigma model and 
in the motion of the D7-branes in the gravity dual side 
can be written as a version of the out-of-time-ordered correlators.

As briefly described in the main part of the paper, recent development in quantum chaos
and AdS/CFT is based on the so-called out-of-time-ordered correlator
\begin{align}
C(t) = - \langle [W(t), V(0)]^2\rangle \, .
\label{oto}
\end{align}
The significance of this correlator was emphasized in \cite{Maldacena:2015waa} which ignited
important subsequent developments in quantum chaos and AdS/CFT.
The important conjecture of \cite{Maldacena:2015waa} is the upper bound for the Lyapunov exponent
\begin{align}
L \leq \frac{2\pi T}{\hbar} \, ,
\label{bou}
\end{align}
for the out-of-time-ordered correlator \eqref{oto} which is properly regularized.
In this supplemental material
we provide how our Lyapunov exponent calculated in the gravity dual can be written as
a version of this out-of-time-ordered correlator.

\subsection{Relation between classical chaos and quantum correlator}

Before a presentation of the precise relation between our results and the out-os-time-ordered correlators,
it is instructive to have a brief review of how \eqref{oto} can be related to a classical chaos
in the semiclassical limit \cite{Maldacena:2015waa} based on the argument in \cite{Larkin}.
Consider a classical particle motion where the location of the particle at time $t$ is given by $q(t)$. 
Its conjugate operator is $p(t)$. The classical chaos means the exponential growth
of a tiny difference at the initial conditions, 
\begin{align}
\left(\frac{\delta q(t)}{\delta q(0)}\right)^2 \sim \exp [2L t] \, .
\end{align}
Here $\delta q(0)$ is the initial difference of the position of the particle. The growth rate is
given by the Lyapunov exponent $L$, and the factor $2$ (and the square in the left hand side) 
is just for a later purpose. The left hand side can be re-written by a Poisson bracket, 
\begin{align}
\left(\{ q(t), p(0)\}_P\right)^2 \sim \exp [2L t] \, .
\end{align}
Now, when one goes to a quantum mechanical regime, the Poisson bracket is replaced by a
commutator $[\hat{q}(t), \hat{p}(0)]$. 
Therefore, the out-of-time-ordered correlator in quantum mechanics
can capture the chaos,
\begin{align}
- \left\langle 
 [\hat{q}(t), \hat{p}(0)]^2
\right\rangle
 \sim \hbar^2 \exp [2L t] \, .
\end{align}

To illustrate the relation between the operator commutation and the classical sensitivity, let us consider the
simplest example 
of a harmonic oscillator in 1 dimension. 
Since it is solvable, one can evaluate the operator commutation explicitly.
We start with a harmonic oscillator Hamiltonian
\begin{align}
H = \frac{p^2}{2m} + \frac12 m \omega^2 q^2 \, .
\end{align}
Then it is easy to calculate the relevant commutator of the operators in the Heisenberg representation,
\begin{align}
\frac{1}{ i \hbar}[\hat{q}(t_1), \hat{p}(t_2)] = \cos[\omega(t_2-t_1)] \, .
\label{xp}
\end{align}
We compare this with a classical counterpart. The classical solution is $q(t)=A\cos[\omega t + \alpha]$. So
an infinitesimal deviation at the initial time $t=t_2$ is produced by a slight shift of time while keeping the
total energy of the solution,
\begin{align}
\delta q(t_2) = -A \omega \delta t \sin[\omega t_2 + \alpha] 
\end{align}
The resultant deviation at final time $t=t_1 (>t_2)$ is 
\begin{align}
\delta q(t_1) = -A \omega \delta t \sin[\omega t_1 + \alpha] 
\end{align}
So, altogether, the classical growth ratio is
\begin{align}
\frac{\delta q(t_1)}{\delta q(t_2)} 
&= \frac{\sin[\omega t_1 + \alpha]}{\sin[\omega t_2 + \alpha]} 
\nonumber \\
&= \sin[\omega (t_2-t_1)] \cot[\omega t_2 + \alpha] + \cos[\omega(t_2-t_1)] \, .
\label{dxdx}
\end{align}
Now, let us take an average over the initial condition. The averaging is about $\alpha$ for this case of the 
harmonic oscillator. Then
\begin{align}
\int_0^{2\pi} \frac{d\alpha}{2\pi}
\; \frac{\delta q(t_1)}{\delta q(t_2)} =\cos[\omega(t_2-t_1)] \, .
\end{align}
The result is identical to the quantum result \eqref{xp}. Here we learn that to be related to the Heisenberg quantum picture
one needs an integration over the initial conditions at the classical side. 

Concerning the Lyapunov exponent, in the above example of the harmonic oscillator,
we can take a pure imaginary $\omega = i\tilde{L}$. 
In this inverse harmonic oscillator, a Lyapunov exponent can be computed,
though the system is not really chaotic because the classical motion is not
bounded.
Nevertheless, it would be instructive to see how the exponential growth rate $\tilde{L}$
can be detected by the quantum operator commutator and the classical analysis of the deviation.
The operator equation
at late times is now replaced by
\begin{align}
\frac{1}{ i \hbar}[\hat{q}(t), \hat{p}(0)] \simeq e^{\tilde{L}t} \, ,
\label{xp2}
\end{align}
while the classical evaluation \eqref{dxdx} before the averaging over $\alpha$
(as there is no period in this case) at late times coincides with it,
\begin{align}
\frac{\delta q(t_1)}{\delta q(t_2)} \simeq  e^{\tilde{L}t} \, .
\label{dxdx}
\end{align}
Therefore, again for the inverse harmonic oscillator, we find the coincidence
between the Heisenberg operator commutator and the classical trajectory deviation,
at later times.


\subsection{Classical chaos written by operators}

The out-of-time-ordered correlator
in \cite{Maldacena:2015waa} 
 is evaluated with the thermal ensemble, specified by the temperature $T$. However, normally
for classical chaos one does not introduce the temperature $T$: the growth rate is calculated by
a classical deterministic motion of $q(t)$ for a given set of initial conditions. For the initial conditions,
one needs to provide energy since it is normally conserved for Hamiltonian systems. Therefore
the classical Lyapunov exponent is a function of energy, rather than the temperature. 
From this argument, we notice that the classical chaos is normally given by a micro canonical ensemble
(which is specified by an energy) while the chaos measured by \eqref{oto} in \cite{Maldacena:2015waa}
is given by a canonical ensemble (which is specified by a temperature).

Crudely speaking, the temperature $T$ determines the rough energy scale of excited states, so one could
say that the energy $\sim T$. More precisely, the micro canonical ensemble can be integrated to provide the canonical ensemble
by statistically averaging physical observables with the weight $\exp[-\beta H]$.

With this difference in mind, we describe how a Lyapunov exponents measured in a classically deterministic system
can be written as a version of the out-of-time-ordered correlator. 
According to the correspondence described above, the classical Lyapunov exponent
can be described as
\begin{align}
- \left\langle E,{\cal M} |
 [q(t), p(0)]^2
 | E, {\cal M} 
\right\rangle
 \sim \hbar^2 \exp [2L t] \, .
 \label{em}
\end{align}
In other words, 
\begin{align}
L = \lim_{t\to\infty}
\frac1{2t} \log
\biggm|
\left\langle E, {\cal M} |
 [q(t), p(0)]^2
 | E, {\cal M} 
\right\rangle
\biggm|
 \, .
 \label{otoss}
\end{align}
Here, note that the state is specified by the classical energy $E$ and the initial condition ${\cal M}$. In general, for 
a given energy $E$, there are many degenerate states. This index ${\cal M}$ picks up one state among
those energy-degenerate states. The existence of a chaos, that is, a sensitivity to the initial conditions, is 
related to this index ${\cal M}$.\footnote{Note that the classical energy in the $\hbar\to 0$ limit has a large quantum degeneracy. For a fixed classical energy
$E$, the quantum energy levels $E_k$ which approach $E$ in the $\hbar\to 0$ limit is in the range $E-n\hbar < E_k <
E+n\hbar$
for some finite $n$. In the classical limit, this finite number $n$ can be arbitrarily large, which provides the
infinite variety of 
classical initial conditions for a given energy $E$. The moduli ${\cal M}$ in \eqref{otoss} needs to be understood
as this degeneracy. For the harmonic oscillator, any given quantum energy level $E_k$, the state is not degenerate.
However, for a given classical energy $E$, there are infinitely many quantum states, and superposition of all of them
gives rise to the classical initial condition $\alpha$.
}

As described earlier, our out-of-time-ordered correlator is evaluated by $|E,{\cal M}\rangle$, which is 
a state in a micro canonical ensemble. To make a connection to the thermal ensemble, we need to sum over
all states with the canonical weight $\exp[-\beta E V]$ where $E$ is the energy density and $V$ is the spatial volume of the
system. The thermal sum of the expression \eqref{em} results in
\begin{align}
\int \! dE \int \! d{\cal M}(E) \exp[-\beta E V]
\left\langle E,{\cal M} |
 [q(t), p(0)]^2
 | E, {\cal M} 
\right\rangle
\end{align}
where $d{\cal M}(E)$ integral normally gives the density of states. The classical Lyapunov exponent $L$ in \eqref{em} 
depends on $E$ and ${\cal M}$, so substituting \eqref{em} into this expression shows
\begin{align}
\int \! dE \int \! d{\cal M}(E) \, \exp[-\beta EV]
\exp [2L(E,{\cal M}) t] \, .
\end{align}
So, the thermal Lyapunov exponent $L_T$ studied in \cite{Maldacena:2015waa} could be related to the
classical  Lyapunov exponent
$L(E,{\cal M})$ as
\begin{align}
L_T = \lim_{t\to\infty}
\frac1{2t} \log
\left[
\int \! dE \int \! d{\cal M}(E) 
\right.
\hspace{25mm}
\nonumber 
\\
\, \exp[-\beta EV]
\exp [2L(E,{\cal M}) t]
\biggm]
 \, .
 \label{rel}
\end{align}
This is the relation between the Lyapunov exponent $L(E,{\cal M})$ of our classical evaluation and the thermal
Lyapunov exponent $L_T$ given by the 
out-of-time-ordered correlator.
In general, there is no simple analytic expression for the explicit relation.\footnote{
If $L(E)$ grows linearly in $E$, the thermal weight $\exp[-\beta EV]$ cannot
make the energy integral convergent, and the integral itself is not well-defined. Therefore, the powers in $E$ appearing in
the classical Lyapunov exponent
seems to be bounded from above: $c<1$ for $L(E)\sim E^c$.
This condition may be regarded as a classical version of the bound \eqref{bou}.
}


\subsection{Chaos of chiral condensate in operator language}

Let us start with the $\sigma$ model  example. The $\sigma$ model consists of the meson fields $\sigma$
and $\pi$, and their conjugate operators are given by 
\begin{align}
P_\sigma = \dot{\sigma} \,, 
\quad
P_\pi = \dot{\pi} \, .
\end{align}
There are various ways to define the ``distance" between two orbits in the phase space of the $\sigma$ model,
but we adopted simply the distance in $\sigma(t)$ for given two initial conditions, because we are interested 
in the chiral condensates. Then, our Lyapunov exponent for the chiral condensate is written as
\begin{align}
L = \lim_{t\to\infty}
\frac1{2t} \log
\biggm|
\left\langle E, {\cal M} |
 [\sigma(t), \dot{\sigma}(0)]^2
 | E, {\cal M} 
\right\rangle
\biggm|
 \, .
\end{align}

We can argue an evaluation of \eqref{rel} once a model is given.
In the case of our $\sigma$ model, as we have observed in figure.~3, the Lyapunov exponent is 
nonzero only for a limited region in the energy scan. So, let us simply assume that the region is
very narrow and the Lyapunov exponent is nonzero only at a certain energy density $E=E_0$.
Then effectively the energy integration in the right hand side of \eqref{rel} is evaluated for 
the limited region (a delta function  $\delta(E-E_0)$),
and we obtain
\begin{align}
L_T  = L(E=E_0)\, .
\label{LL}
\end{align}
So the thermal Lyapunov exponent is equal to our classical Lyapunov exponent at the region.

Let us discuss the meaning of \eqref{LL} in view of the bound \eqref{bou}. 
Although the bound \eqref{bou} for $L_T$ is for a properly regularized
four-point function which is different from $[\sigma(t),\dot{\sigma}(0)]^2$ used above, it may be instructive to discuss
how the bound could be consistent \eqref{LL}. It appears that \eqref{LL} contradicts with the bound \eqref{bou} for a small 
temperature. But notice that a possible ``violation" needs a very small temperature comparable to ${\cal O}(\hbar)$
and thus the classical evaluation of the Lyapunov exponent is not validated. 
Let us argue this a bit more in detail.
The quantum time evolution is by a unitary operator $\exp[iHt/\hbar]$, thus the validity of the classical picture 
is for $Ht \gg \hbar$. We are looking at the Lyapunov growth, so the time scale is $t \sim 1/L$. Using this,
the classical analysis needs $H \gg \hbar L$. Suppose the Lyapunov exponent is constant.
Then a violation of the bound \eqref{bou} needs a low temperature $\hbar L > 2\pi T$. Together with the
classical validity, we obtain $H \gg 2\pi T$. This means that the thermal weight $\exp[-\beta H]$ is highly 
suppressed for this temperature, so it
cannot appear in perturbation theory: the thermal sum in \eqref{rel} at such a low temperature could be overwhelmed 
by other perturbative corrections. Therefore, the ``violation" is not an immediate conclusion, we need more careful 
quantum treatment
of the evaluation of the thermal partitions.


In this paper we considered a homogeneous time-dependent field configuration of the mesons, but in general
the mesons are inhomogeneous. The operators in the expression \eqref{otoss} is naturally evaluated
at the same position in the three-dimensional space. In QCD chiral perturbation, there are some study of
homogeneous meson configurations, which is called $\delta$-regime
\cite{Hasenfratz:2009mp}. 
In this $\delta$-regime, the volume $V$ is taken to zero. However, $H=EV$ cannot be so small
if we rely on classical treatment, as mentioned above: $H =EV\gg \hbar L$ is necessary.

Let us briefly discuss the out-of-time-ordered correlator for 
our main example of the ${\cal N}=2$ supersymmetric QCD. The discussion is the same except for the
difference of the operators. The operators for which we measured the Lyapunov exponent are $x\sim 
\bar{q}^1q^2+\bar{q}^2q^1$
and $y \sim \bar{q}^1\gamma_5 q^2-\bar{q}^2\gamma_5 q^1$. The Lyapunov exponent is defined in a similar manner,
\begin{align}
L = \lim_{t\to\infty}
\frac1{2t} \log
\biggm|
\left\langle E, {\cal M} |
 [x(t), \dot{x}(0)]^2
 | E, {\cal M} 
\right\rangle
\biggm|
 \, .
\label{otoxx}
\end{align}
Our numerical analysis given in the right panel of Fig.~5 shows that for fixed $\lambda$ and $N_c$ the
Lyapunov exponent grows as we increase the energy density $E$. 
Evaluation of the thermal Lyapunov exponent
through \eqref{rel} needs an analytic 
functional form of $L(E)$ and also the density of state. We would like to leave it for our future study.


\end{document}